\documentclass[onecolumn,16pt,GungsuhChe]{revtex4}%16pt,GungsuhChe
\usepackage{amsmath,graphicx,graphics,amsmath,CJK,color}
\usepackage{multirow,fancyhdr,color,bm,tabularx,psfrag}
\usepackage[dvipdfm,colorlinks=true,linkcolor=blue,urlcolor=blue,citecolor=blue]{hyperref}
\def\headrule{\kern 1mm \hrule width 20cm \kern -1mm}%
\def\footnoterule{\kern 1mm \hrule width 7cm \kern 2.2mm}%
\def\textwidth{16cm}\oddsidemargin 0.1cm\evensidemargin 0.1cm
\begin{document}

\title{\LARGE Wider frequency domain for negative refraction index in a quantized composite right-left handed transmission line}

%\thanks{}

\author{Qi-Xuan Wu}
\affiliation{Faculty of Foreign Languages and culture, Kunming University of Science and Technology, Kunming, 650500, PR China}
\author{Shun-Cai Zhao}
\email[Corresponding author: ]{ zhaosc@kmust.edu.cn }
\affiliation{Department of Physics, Faculty of Science, Kunming University of Science and Technology, Kunming, 650500, PR China}
\date{\today}

\begin{abstract}
The refraction index of the quantized lossy composite right-Left handed transmission line (CRLH-TL) was deduced in the thermal coherence state. The results show negative refraction index (herein the left-handedness) can be implemented by the electric circuit dissipative factors(i.e., the resistances \(R\) and conductances \( G\)) in the higher frequency bands (1.446GHz\(\leq\omega\leq \) 15GHz), and flexibly adjusted by the left-handed circuit components (\(C_l\), \(L_l\)) and the right-handed circuit components (\(C_r\), \(L_r\)) at a lower frequency (\(\omega\)=0.995GHz) . The flexible adjustment for left-handedness in a more wider bandwidth will be significant to the microscale circuit design for the CRLH-TL and may pave the theoretical preparation for its compact applications.
\begin{description}
\item[PACS: ]{81.05.Xj, 78.67.Pt, 78.20.Ci, 42.50.Gy }
\item[Keywords:]{ Quantized composite right-Left handed transmission line; negative refraction index;frequency domain}
\end{description}
\end{abstract}

\maketitle
A well-established route for constructing negative refractive index materials (NRM)
is based on Veselago`s theory\cite{1} of left-handed materials (LHM), simultaneous
negative permittivity (\(\epsilon\)) and magnetic permeability (\(\mu\)) with different
types of metamaterials\cite{2,3,4,5,6,7}. Although very exciting from a physics point of view, the negative
\(\epsilon\) and \(\mu\) produced by electromagnetic resonance may bring about a very high loss\cite{8,9}
and narrow bandwidth consequently. Due to the weaknesses of resonant-type structures, three groups almost
simultaneously introduced a transmission line (TL) approach for NRM\cite{10,11,12,13}, i.e., the composite
right-Left handed transmission line (CRLH-TL), which refers to the right-handedness accompanying the positive refraction index
at high frequencies and to the left-handedness with the negative refraction index (NRI ) at lower frequencies\cite{a}. CRLH-TL initially the non-resonant-type one, is perhaps one of the most representative and potential candidates due to its low loss, broad operating frequency band,
as well as planar configuration\cite{14,a}, which is often related with easy fabrication for NRI
applications in a suite of novel guided-wave\cite{16}, radiated-wave\cite{17}, and refracted-wave
devices and structures\cite{18,19}. Nowadays, the CRLH-TLs show a tendency to the compact applications influenced by the nanotechnology and microelectronics\cite{b,c,d}. Recently, a new class of miniaturized nonreciprocal leaky-wave antenna is proposed for miniaturization, nonreciprocal properties and wide-angle scanning at the same time\cite{20}. With four unit cells of CRLH-TL a wide-band loop antenna is proposed in a compact size\cite{21}.

However, when the compact size of the CRLH-TL approaches the Fermi wavelength, the quantum effects\cite{22,23,24} on the CRLH-TL must be taken into account. In our former work, we firstly deduced the quantum features of NRI of the lossless mesoscopic left-handed transmission line (LH-TL)\cite{25}. Then we quantized the lossy LH-TL\cite{26} and we discussed the quantum influence of dissipation on the NRI\cite{26} in a displaced squeezed Fock state. And some novel quantum effects were revealed and were reputed significance for the miniaturization application of LH-TL.

In this paper, a flexible adjustment for negative refraction index from a wider frequency bandwidth in the quantized lossy CRLH-TL is achieved in the thermal coherence state. And the paper is organized as follows. In Sec.2, we quantize the travelling current field in the unit-cell circuit of the CRLH-TL, and deduce its refraction index in the thermal coherence state. Then we evaluate the refraction index via the numerical approach in Sec.3. Sec.4 presents our summary and conclusions.
Section 5 presents our summary and conclusions.

\section{The quantized refraction index in the thermal coherence state }

\begin{figure}[htp]
\center
\includegraphics[totalheight=1.8 in]{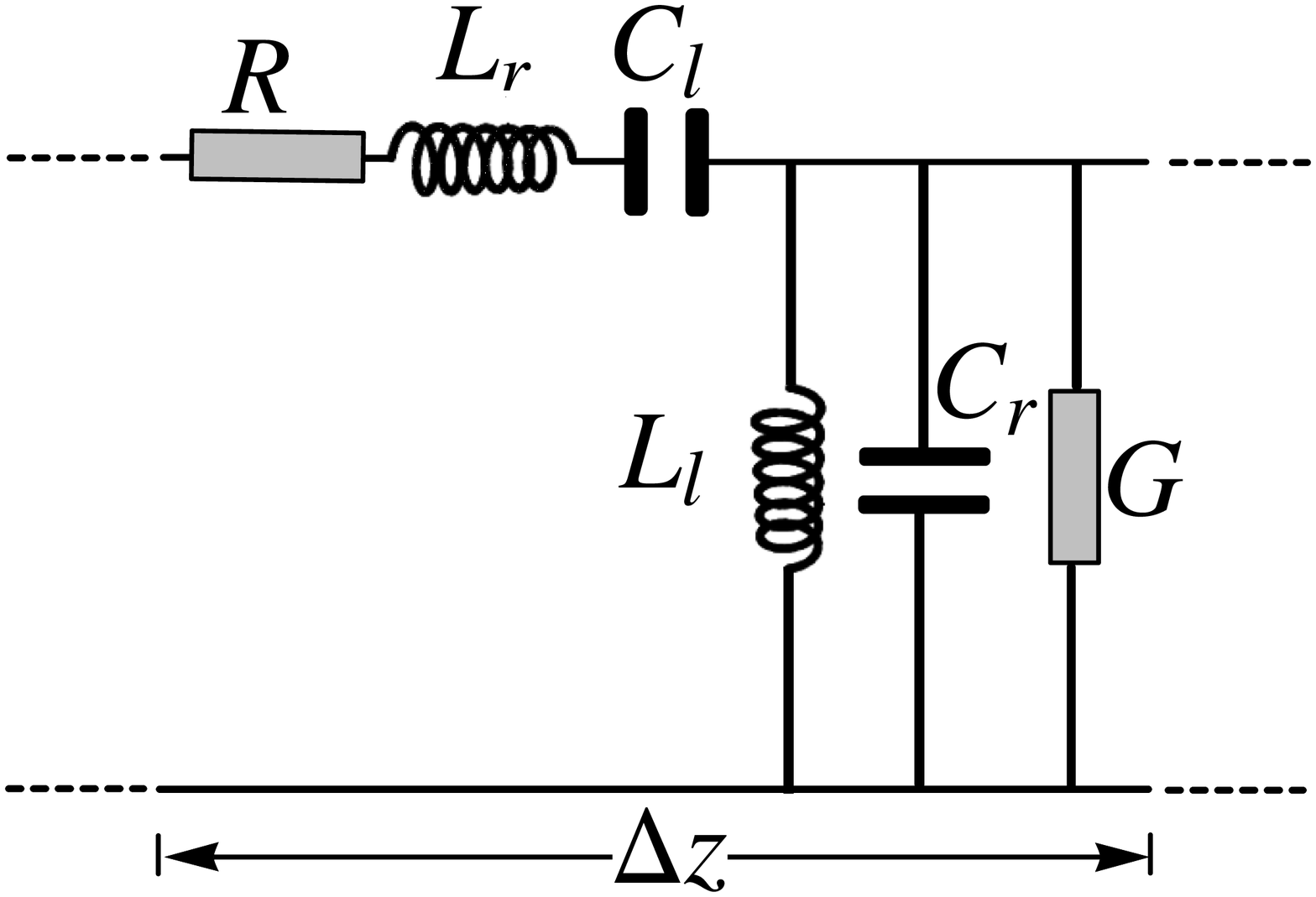 }
\caption{chematic diagram of an equivalent unit-cell circuit of the mesoscopic lossy CRLH-TL.}
\end{figure}\label{Fig.1}
The equivalent unit-cell circuit model of the proposed lossy CRLH-TL is shown in Fig. 1. Comparing to the lossless CRLH-TL\cite{27}, the imported resistance \(R\) and conductance \( G\) represent the loss except the series capacitor \(C_l\) and inductance \(L_r\), shunt inductance \(L_l\) and capacitor \(C_r\)\cite{28} in each unit cell circuit. And the dimension \(\Delta z\) of the equivalent unit-cell circuit is much less than the wavelength at operating frequency. Hence, let us now consider Kirchhoff¡¯s voltage and current laws for this unit-cell circuit in Fig. 1, which respectively read
\begin{align}
u(z)=j(z)[R+\frac{1}{i\omega C_l}+i\omega L_R]\Delta{z}+u(z+\Delta{z}),\\
j(z)=u(z)[G+\frac{1}{i\omega{L_l}}+i\omega C_R]\Delta{z}+j(z+\Delta{z})
\end{align}
where \(u(z)\) is the voltage, \(j(z)\) is the current, and \(\omega\) is the angle frequency. When \(\Delta z\rightarrow 0\),
the above equations, lead to the following system:
\begin{align*}
 \frac{\partial^2{j(z)}}{\partial{z^2}}=[G+\frac{1}{i\omega{L_l}}+i\omega C_R]\times[R+\frac{1}{i\omega C_l}+i\omega L_R]j(z)
\end{align*}
The above electric current equation leads to the forward plane-wave solution:
\begin{align}
j(z)=\exp{(-\sigma z)}[A e^{(i\beta{z})} + A^* e^{(-i\beta{z})}]
\end{align}

\noindent in which
 \begin{align}
\sigma+i\beta=\sqrt{[G+\frac{1}{i\omega{L_l}}+i\omega C_R]\times[R+\frac{1}{i\omega C_l}+i\omega L_R]}
\end{align}
In Eq.(3) \( A^{*}\) is the complex conjugate of \(A \) for normalization purposes. We adopt the quantization method similar to Louisell \cite{29} to achieve the current operator. In Fig.1 the given unit-length, i.e.,\(z_0=m\lambda\) where \(\lambda\) is the wavelength labelled typically by wavenumber k and frequency \(\omega\), its Hamiltonian can be written as follows,
\begin{align*}
H&=\frac{1}{2}\int_{0}^{z_{0}}(Rj^{2}+L_{r}j^{2}+C_{l}u_{C_{l}}^{2}+\frac{1}{G}j_{G}^{2}+L_{l}j_{L_{l}}^{2}+C_{r}u^{2})dz \\
&=F A^{*}Az_{0}
\end{align*}
\noindent where F=\(\frac{1+R\omega C_{l}+\omega^{2} L_{r} C_{l}}{\omega C_{l}}+ \frac{L_{l}^{2}\omega^{2}( G+1)}{(1+\omega L_{l}G+ \omega^{2}L_{l}C_{r})^{2}}+\omega C_{r}(\frac{1+R\omega C_{l}+\omega^{2} L_{r} C_{l}}{\omega C_{l}}+\frac{\omega L_{l}}{1+\omega L_{l}G+ \omega^{2}L_{l}C_{r}})^{2}\). Then, with the definitions \(A=a\sqrt{\frac{\hbar\omega}{F z_{0}}}, A^{*}=a^{*}\sqrt{\frac{\hbar\omega}{F z_{0}}}\) we achieve the Hamiltonian of the unit-cell circuit,
\begin{align}
H=\hbar\omega a^{*}a
\end{align}
According to the canonical quantization principle, we can quantize the system by operators \(\hat{q}\) and \(\hat{p}\), which satisfy the
commutation relation \( [\hat{q},\hat{p}]\)= i\(\hbar\). The annihilation and creation operators \(\hat{a}\) and \(\hat{a}^{+}\)
are defined by the relations
\begin{equation*}
\begin{split}
\hat{a}=\frac{1}{\sqrt{2\hbar\omega}}(\omega\hat{q}+i\hat{p}),\qquad \hat{a}^\dag=\frac{1}{\sqrt{2\hbar\omega}}(\omega\hat{q}-i\hat{p})
\end{split}
\end{equation*}
Thus the quantum Hamiltonian of Fig.1 can be rewritten as \( \hat{H}=\hbar\omega{\hat{a}^\dag\hat{a}}=\frac{1}{2}(\omega^{2}\hat{q}^{2}+\hat{p}^{2})
\). Thus the current in the lossy unit cell equivalent circuit for CRLH-TL can be quantized as
\begin{align}
j(z)=\sqrt{\frac{\hbar\omega}{F z_{0}}}\exp{(-\sigma z)}[\hat{a} e^{(i\beta{z})} + \hat{a}^\dag e^{(-i\beta{z})}]
\end{align}
The thermal coherent state we adopted here is \(|\alpha\rangle=D(\alpha)|0\tilde{0}\rangle_{T}=D(\alpha)T(\theta)|0\tilde{0}\rangle\)\cite{30,31,32}.
Where \(D(\alpha)=\exp(\alpha a^{\dag}-\alpha^{*}a)\), \(T(\theta)=\exp[-\theta(a\tilde{a}-a^{\dag}\tilde{a}^{\dag})]\) and with \(\sinh^{2}(\theta)=[\exp(\frac{\hbar\omega}{k_{B}T})-1]^{-1} \), in which \(k_B\) is the Boltzmann constant. In thermo-field dynamics (TFD) theory, the tilde space accompanies with the Hilbert space, and the tilde operators commute with the non-tilde operators\cite{33}.
Thus the creation and annihilation operators \(\hat{a}^{\dag}\), \(\hat{a}\) associate with their tilde operators \(\tilde{\hat{a}}^{\dag}\), \(\tilde{\hat{a}}\) according the rules\cite{33}, \([\tilde{\hat{a}},\tilde{\hat{a}}^{\dag}]=1\),\([\tilde{\hat{a}},\hat{a}]=[\tilde{\hat{a}},\hat{a}^{\dag}]=[\hat{a},\tilde{\hat{a}}^{\dag}]=0\).
And we can prove the following equalities easily,
\begin{equation}
\begin{split}
T^{\dag}(\theta)aT(\theta)=ua+v\tilde{a}^{\dag}  ,
\qquad
T^{\dag}(\theta)a^{\dag}T(\theta)=ua^{\dag}+v\tilde{a} ,
\qquad    \\
D^{\dag}(\alpha)a D(\alpha)=a+\alpha    ,
\qquad
D^{\dag}(\alpha)a^{\dag}D(\alpha)=a^{\dag}+\alpha^{*}  ,
\qquad    \\
\tilde{D}^{\dag}(\tilde{\alpha})\tilde{a} \tilde{D}(\tilde{\alpha})=\tilde{a}+\tilde{\alpha},
\qquad
\tilde{D}^{\dag}(\alpha)\tilde{a}^{\dag}\tilde{D}(\alpha)=\tilde{a}^{\dag}+\tilde{\alpha}^{*}
\qquad \\
\end{split}
\end{equation}
with \(u=\cosh(\theta)\), \(v=\sinh(\theta)\). Then from Eq.(6) to Eq.(7), the quantum fluctuation of the current in the thermal coherent state is
\begin{align}
\langle(\Delta \hat{j})^{2}\rangle=2u^{2}\frac{\hbar\omega}{Fz_{0}}\exp(-2\sigma z)
\end{align}
It notes that \( \sigma z\) is infinitesimal when \(z\) is a dimensionless, and we can transform Eq.(8) by using the first order approximation of Taylor expansion as follow,
\begin{align}
\frac{\langle(\Delta \hat{j})^{2}\rangle Fz_{0}}{2u^{2}\hbar\omega}=1-2\sigma z
\end{align}
With the relation \(2\sigma\beta=\eta\) deduced from Eq.(4), and the relation (  \( n=\frac{c_0 \beta}{\omega}\)\cite{34},\(c_0\) is the light speed in vacuum) between propagation constant and refraction index in the CRLH-TL, the refraction index can be deduced by Eq.(9),

\begin{align}
n=\frac{2\hbar c_0 u^{2} \eta z}{2u^{2}\hbar\omega-\langle(\Delta \hat{j})^{2}\rangle F z_{0}}
\end{align}

\section{Numerical simulations and discussions}

\begin{table*}[t]
\footnotesize
\caption{Parameters of the circuit components in the quantized lossy CRLH-TL.}\label{tab1}
\tabcolsep 15pt %space between two columns. ????????§Þ???
\begin{tabular*}{\textwidth}{c|c|c|c|c|c|c|c}
\hline
         & \(C_l (pF)\) & \(L_l (pH) \)  & \(C_r (nF)\)  & \(L_r (nH)\)  & \(\omega(GHz) \)  & \(R (\mu\Omega) \)  &  \(G (\mu S) \) \\
\hline
Fig.2(a) & 3.667        & 416.7          & 9.80          & 129.8         &  ---              & ---                 &  99.8  \\
Fig.2(b) & 3.667        & 416.7          & 9.80          & 129.8         &  ---              & \(0.95\)            &  ---   \\
Fig.3(a) & ---          & 416.7          & ---           & 129.8         & 0.995             & \(0.35 \)           &  99.8  \\
Fig.3(b) & 1.000        & 416.7          & ---           & ---           & 0.995             & \(0.35 \)           &  99.8  \\
Fig.4(a) & ---          & 416.7          & 98.0          & ---           & 0.995             & \(0.35 \)           &  99.8  \\
Fig.4(b) & 285          & ---            & 98.0          & ---           & 0.995             & \(0.35 \)           &  99.8  \\
\hline
\end{tabular*}
\end{table*}

In the previous work\cite{a}, the negative refraction index happens in the lower frequency bandwidth of microwave wave. In order to investigate minus refraction index, we should work with the refraction index Exp.(10). However, the analytical Exp.(10) of the refraction index is rather cumbersome to obtain, so, we follow the numerical approach to analyze the refraction index of the CRLH-TL. And several parameters should be selected before the analysis. The length of the unit CRLH-TL circuit is \(z=4z_{0}=4nm\), and the quantum fluctuation of the current \(\langle(\Delta j)^{2}\rangle=10^{-9}\). Other parameters used in our simulation are listed in Table 1 whose order of magnitudes are referenced from Ref.\cite{35}.

\begin{figure}[htp]
\center
\includegraphics[totalheight=1.8 in]{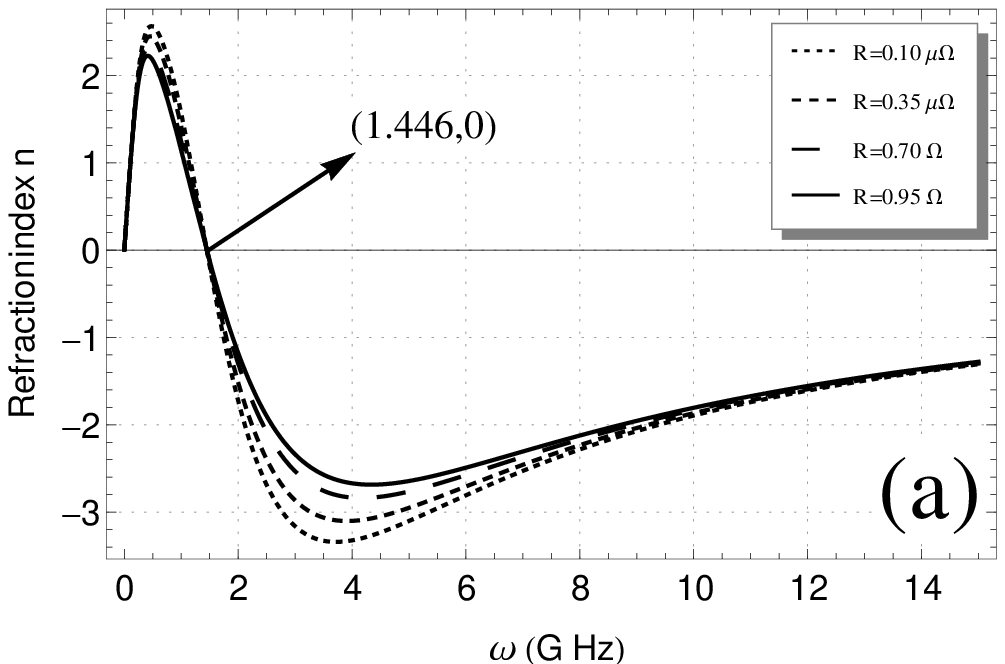 }\includegraphics[totalheight=1.8 in]{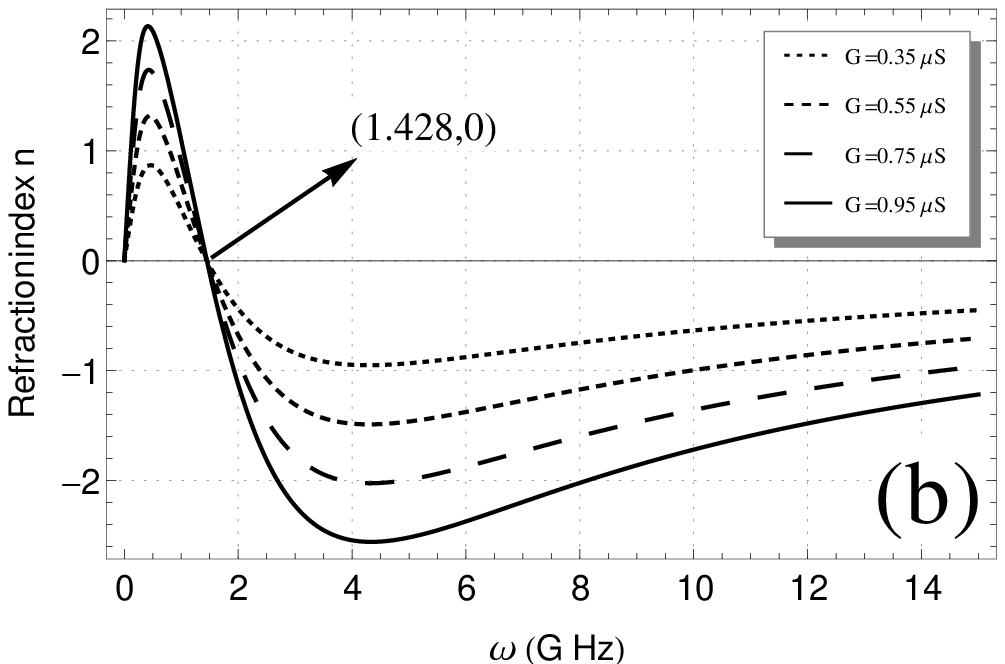 }
\caption{The refraction index $ n $ as a function of the frequency \(\omega\) tuned by different resistances \(R\) in (a) and conductances \( G\) in (b), respectively.}
\end{figure}\label{Fig.2}

The frequency bandwidth for negative refraction index (i.e., the left-handedness ) is of interest for the mesoscopic CRLH-TL. Not only that, the role of the imported resistance \(R\) and conductance \( G\) representing the loss is also what we care about. Fig.2 shows the refraction index dependent the imported resistance \(R\) and conductance \( G\) in the frequency domain. The relation between the refraction index and the resistance \(R\) is shown in Fig.2(a). It notes that in the lower frequency bands [0, 1.446GHz] the refraction indexes are positive, and that its values are negative in the higher frequency bands [1.446GHz, 15GHz] which surpasses the frequency bands for left-handedness mentioned by Ref\cite{a}. And we notice that the resistances \(R\) play a passive role both on the left-handedness and right-handedness frequency bands, i.e. the resistances \(R\) restrain the growth of the refraction index $ n $ in the two frequency bands. Fig.2(b) shows the similar behavior of the conductance \( G\) on the refraction index. The positive refraction indexes in the lower frequency bands [0, 1.428GHz] and the negative refraction index in the higher frequency bands [1.428GHz, 15GHz] can be observed by the curves in Fig.2(b).
While the increasing conductance \( G\) by 0.2 \(\mu S\) can enhance the refraction index from the dotted curve to the solid curve in both the frequency bands, i.e. [0, 1.428GHz] and [1.428GHz, 15GHz]. And we note that the frequency bands for negative refraction index (i.e., the left-handedness ) in Fig.2 surpass which mentioned in Ref.\cite{a}.
In generally speaking, the resistance \(R\) and the conductance \( G\) are not a desired role in electricity out of the Joule heat via the Ohm Law. In this quantized CRLH-TL, the resistance \(R\) and the conductance \( G\) can adjust the refraction index except the classic Ohm Law. Throughout the full Fig.2, this quantized CRLH-TL demonstrates the positive refraction index (i.e., the right-handedness) in the lower frequency bands and the negative refraction index (i.e., the left-handedness) in the higher frequency bands.

A notable question is whether it is possible to realize negative refraction index (i.e., the left-handedness) in the lower frequency bands ? Fig.3 and Fig.4 provide the refraction index dependent the parameters of the circuit components at a lower frequency \(\omega\)=0.995GHz.

\begin{figure}[htp]
\center
\includegraphics[totalheight=1.8 in]{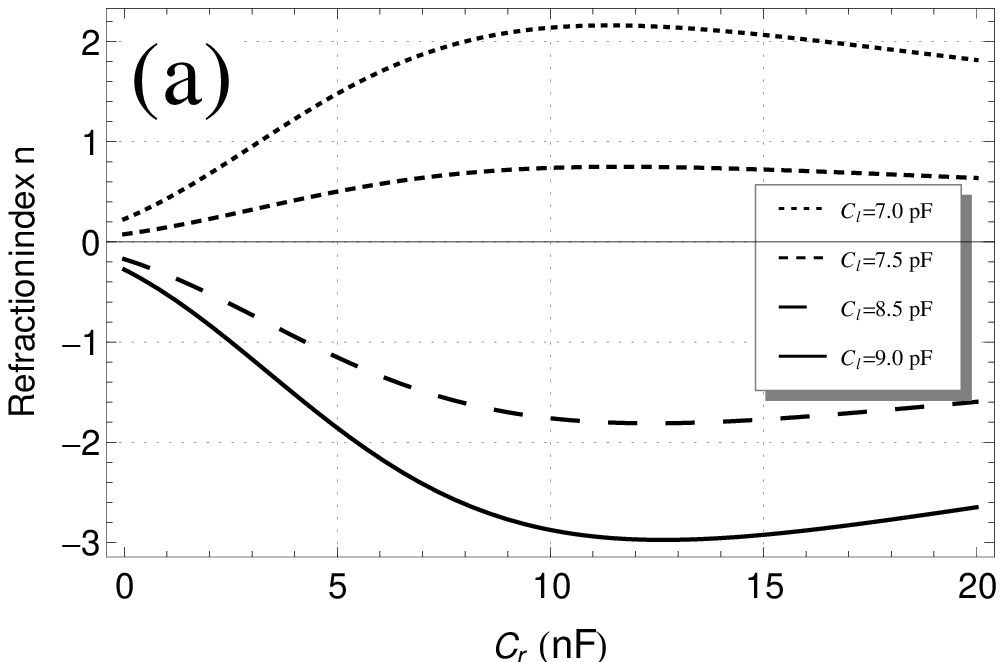 }\includegraphics[totalheight=1.8 in]{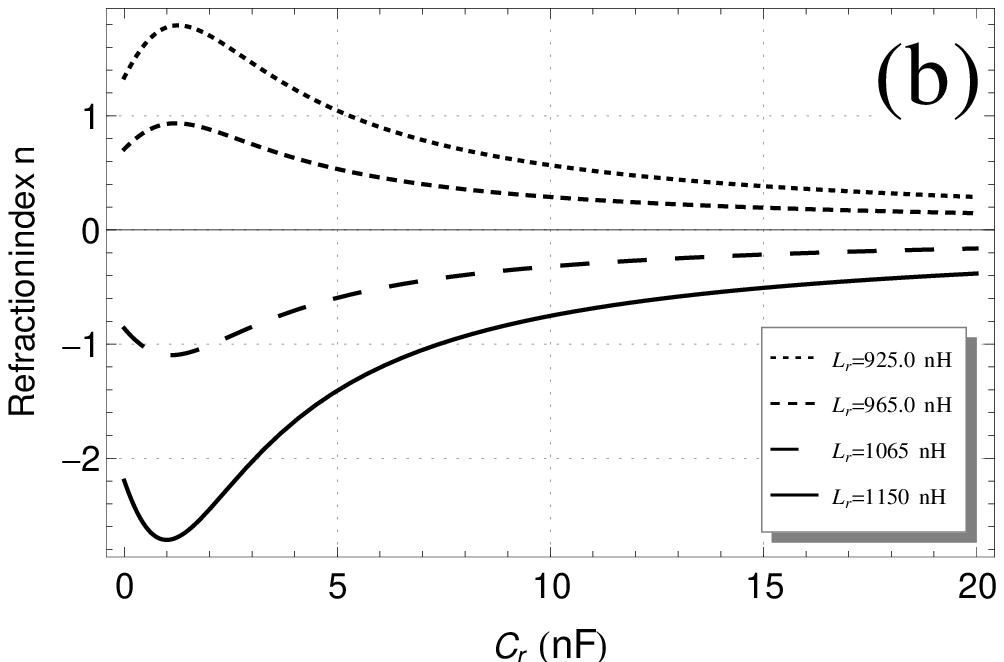 }
\caption{The refraction index $ n $ as a function of the shunt capacitor  \(C_r\) tuned by different series capacitors \(C_l\) in (a) and inductances \(L_r\) in (b), respectively}
\end{figure}\label{Fig.3}

The refraction index dependent the shunt capacitor \(C_r\) is provided in Fig.3 at frequency \(\omega\)=0.995GHz. The curves show that the refraction index is positive when the series capacitor \(C_l\) is assigned by 7.0 \(pF\) and 7.5 \(pF\) in Fig.3 (a). But the refraction index is negative and reaches its maximum in the ranges [10\(nF\), 15\(nF\)] of the shunt capacitor \(C_r\) when the series capacitor \(C_l\) is set by the larger values, i.e., \(C_l\)=8.5\(pF\), 9.0\(pF\), respectively. The curves from dotted to solid in Fig.3(b) show the similar feature when the refraction index is tuned by the different series inductances \(L_r\). However, we also note that a larger negative refraction index can be implemented by a smaller shunt capacitor \(C_r\)\(\approx\)1\(nF\) but a larger series inductances \(L_r\)=1150\(nH\). Fig.3 demonstrates that the refraction index can be negative at a lower frequency \(\omega\)=0.995GHz with the proper parameters of series capacitor \(C_l\) and inductances \(L_r\) within the ranges [0, 20\(nF\)] of the shunt capacitor \(C_r\).

\begin{figure}[htp]
\center
\includegraphics[totalheight=1.8 in]{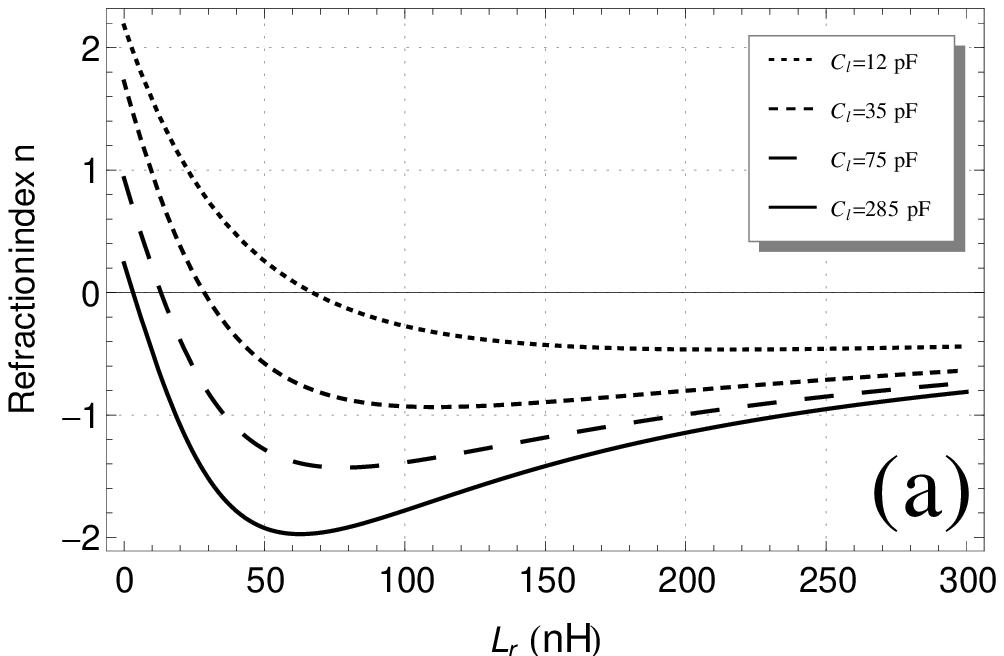 }\includegraphics[totalheight=1.8 in]{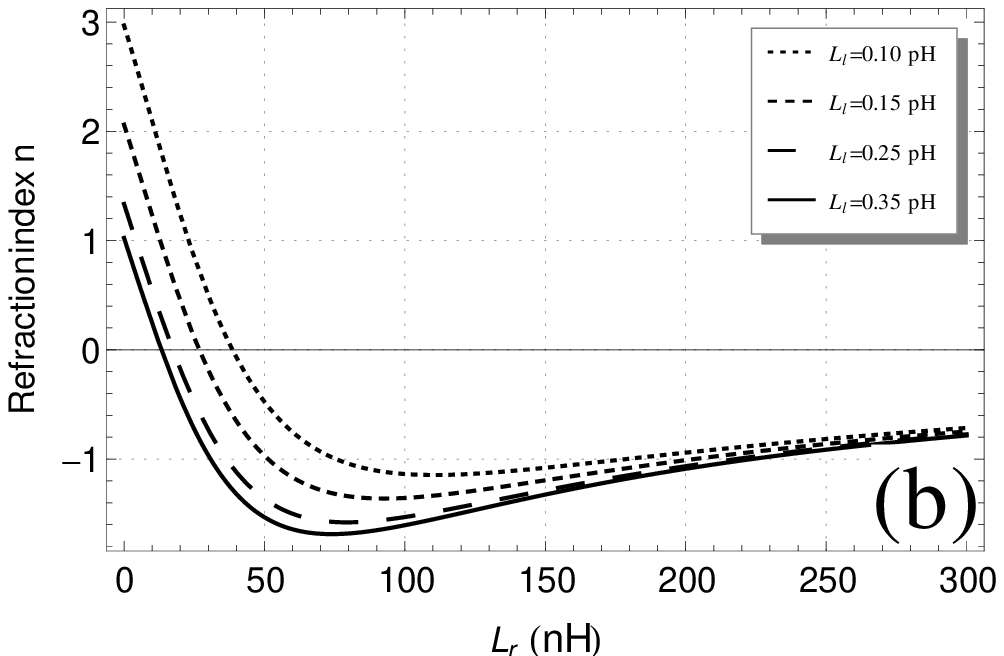 }
\caption{The refraction index $ n $ as a function of the series inductance \(L_r\) tuned by different series capacitors \(C_l\) in (a) and shunt inductances \(L_l\) in (b), respectively.}
\end{figure}\label{Fig.4}

At the lower frequency \(\omega\)=0.995GHz, Fig.4 shows the refraction index dependent the series inductance \(L_r\) tuned by different series capacitors \(C_l\) in (a) and shunt inductances \(L_l\) in (b), respectively. The increasing series capacitors \(C_l\) promote the ranges of the series inductance \(L_r\) for negative refraction index and gradually increase the values of negative refraction index in Fig.4 (a). And the curves in Fig.4 (b) display the similar growth trends for negative refraction index, but the increase of the negative refraction index in Fig.4 (a) is larger than that in Fig.4 (b). These results demonstrate the left-handed circuit components (\(C_l\), \(L_l\)) can enhance the negative refraction index in the mesoscopic lossy CRLH-TL.

Before concluding this paper, we would remark that how to broaden the frequency bands for negative refraction index is an active field for metamaterials,
and the metamaterials within a wider band is always the direction of efforts. However, the frequency bands for negative refraction index only exist in the microwave band introduced by the first researchers\cite{10,11,12,13} in the CRLH-TL. In our current study, we implement the negative refraction index within a wider frequency bands in the quantized CRLH-TL and conclude that the negative refraction index in the higher frequency bands (1.446GHz\(\leq\omega\leq \) 15GHz) tuned by the resistances \(R\) and conductances \( G\) and at a lower frequency (\(\omega\)=0.995GHz) tuned by the parameters of the circuit components. These achievements show the new characteristic for the quantized CRLH-TL and should be paid enough attention in the coming future.

\section{Conclusion}

In present paper, the negative refraction index is implemented within a wider frequency bands in the quantized CRLH-TL. And the frequency domain for negative refraction index
is 1.446GHz\(\leq\omega\leq \) 15GHz with the values for negative refraction index being opposite tuned by the resistances \(R\) and conductances \( G\), respectively.
At a much lower frequency (\(\omega\)=0.995GHz), the negative refraction index can also be flexibly implemented by the left-handed circuit components (\(C_l\), \(L_l\)), and the right-handed circuit components (\(C_r\), \(L_r\)), respectively. The adjusting negative refraction index within a more wider bandwidth in this quantized CRLH-TL is significant to the microscale circuit design and compact applications for the CRLH-TL.

\section{Acknowledgments}
This work is supported by the National Natural Science Foundation of China ( Grant Nos. 61205205 and 6156508508 ),
the General Program of Yunnan Applied Basic Research Project, China
( Grant No. 2016FB009 ) and the Foundation for Personnel training projects of Yunnan Province, China ( Grant No. KKSY201207068 ).

\end{document}